\documentclass{article}
\usepackage{spconf,amsmath,graphicx}

\usepackage{enumitem}
\setlist{nosep, leftmargin=14pt}

\usepackage{mwe} 
\usepackage{hyperref} 

\title{On the benefit of dual-domain denoising in a self-supervised \\low-dose CT setting}
\name{
    \begin{tabular}{c}
        Fabian Wagner$^1$, Mareike Thies$^1$, Laura Pfaff$^1$, Oliver Aust$^2$, Sabrina Pechmann$^3$, Daniela Weidner$^2$,\\ Noah Maul$^1$, Maximilian Rohleder$^1$, Mingxuan Gu$^1$, Jonas Utz$^4$, Felix Denzinger$^1$, Andreas Maier$^1$
    \end{tabular}
}
\address{$^1$Pattern Recognition Lab, FAU Erlangen-Nürnberg, Germany \\
$^2$Department of Rheumatology and Immunology, FAU Erlangen-Nürnberg, Germany \\
$^3$Fraunhofer Institute for Ceramic Technologies and Systems IKTS, Germany \\
$^4$Department AIBE, FAU Erlangen-Nürnberg, Germany}
\begin{document}
%
\maketitle
\begin{abstract}
Computed tomography (CT) is routinely used for three-dimensional non-invasive imaging. Numerous data-driven image denoising algorithms were proposed to restore image quality in low-dose acquisitions. However, considerably less research investigates methods already intervening in the raw detector data due to limited access to suitable projection data or correct reconstruction algorithms. In this work, we present an end-to-end trainable CT reconstruction pipeline that contains denoising operators in both the projection and the image domain and that are optimized simultaneously without requiring ground-truth high-dose CT data. Our experiments demonstrate that including an additional projection denoising operator improved the overall denoising performance by $82.4\text{--}94.1\,\%$/$12.5\text{--}41.7\,\%$ (PSNR/SSIM) on abdomen CT and $1.5\text{--}2.9\,\%$/$0.4\text{--}0.5\,\%$ (PSNR/SSIM) on XRM data relative to the low-dose baseline. We make our entire helical CT reconstruction framework publicly available that contains a raw projection rebinning step to render helical projection data suitable for differentiable fan-beam reconstruction operators and end-to-end learning.
\end{abstract}
\section{Introduction}
\label{sec:intro}

Low-dose computed tomography (CT) denoising aims for reconstructing high-quality volumetric images from CT acquisitions with reduced patient dose. In the last years, conventional CT denoising algorithms were outperformed by methods using neural networks that allow data-driven optimization. Many of these approaches are trained in a supervised fashion, which requires paired low- and high-dose CT data. Recently, multiple self-supervised methods were proposed that can be trained without ground-truth high-dose target data, which greatly simplifies their applicability. One work demonstrates that learning the mapping between reconstructions of two independent sets of projections can be used to train a CT denoising model \cite{hendriksen2020noise2inverse}. 
\begin{figure}[tb]
\centering
\includegraphics[width=\linewidth]{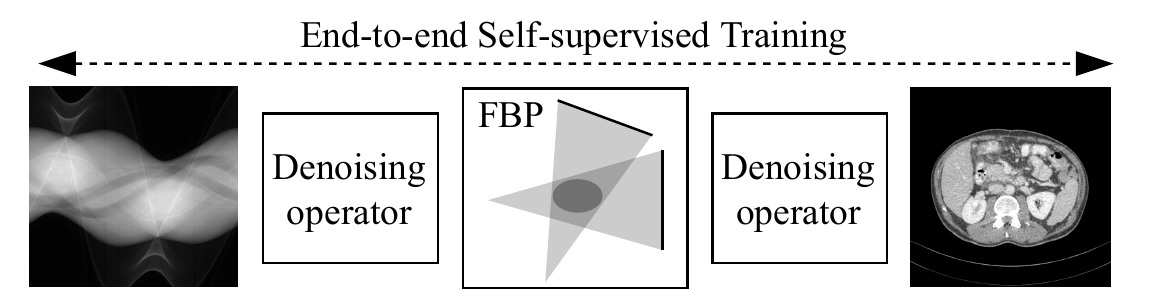}
\caption{Overview of the proposed trainable dual-domain and self-supervised CT denoising pipeline.}
\label{fig:overview}
\end{figure}
Other works employ the similarity of features in neighboring CT slices to learn a mapping to noise-free images \cite{wu2020self}, use the Noise2Noise principle \cite{lehtinen2018noise2noise}, or estimate an underlying noise model to optimize a denoising network \cite{kim2022noise}. All those methods have in common that they perform denoising as a pure post-processing step on reconstructed CT images. However, noise in CT data originates already from the detection process through limited photon statistics and the detector properties itself \cite{yu2012development}. The CT reconstruction algorithms then distribute projection image noise over the entire reconstructed volume, which complicates the noise pattern and denoising task for post-processing algorithms. We believe that there are mainly two reasons why most research focuses on denoising in the image domain. First, projection data can be difficult to handle as it is often acquired on helical trajectories in medical CT scanners. Second, denoising projection images requires a running CT reconstruction algorithm. We selected all $36$ works that reference the most popular public low-dose CT data set (reference \cite{moen2021low}, Google Scholar, Oct 2022) and perform CT data processing. We found that only four of them use the provided raw projection data. All other works only perform experiments starting from the reconstructed images.\\
\begin{figure*}[htb]
\centering
\includegraphics[width=\linewidth]{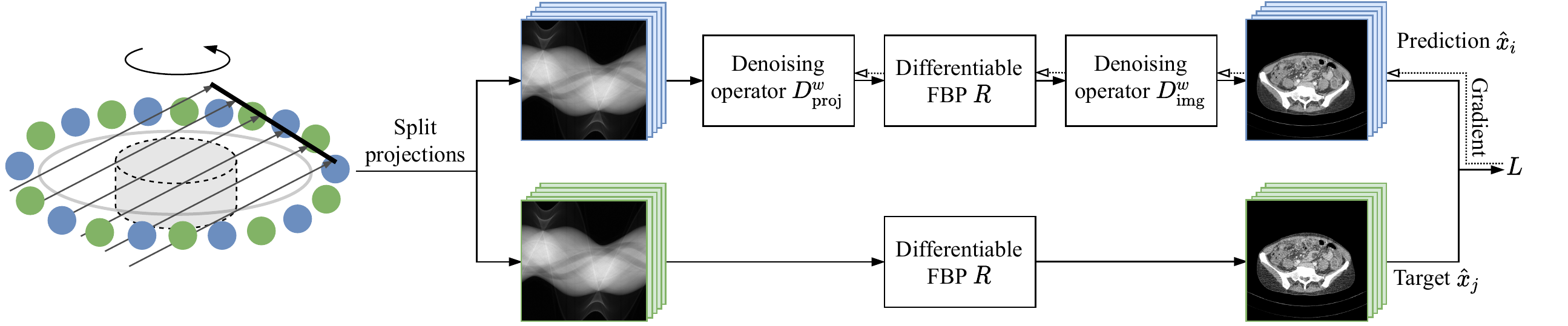}
\caption{Illustration of the proposed end-to-end self-supervised denoising pipeline. Projections are split in two stacks and reconstructed separately. The loss calculated between denoised prediction and reconstructed target stack is backpropagated through $\mathbf{D}_{\text{img}}^w$ and $\mathbf{R}$ to $\mathbf{D}_{\text{proj}}^w$ to optimize all denoising operators. The gradient flow from Eq.~\ref{eq:gradient_flow} is indicated by dashed lines.}
\label{fig:reco_pipeline}
\end{figure*}\noindent
Only a few other works propose denoising CT projection data alone or in combination with the reconstructed images \cite{kim2020unsupervised,wagner2022ultra,ge2022ddpnet,patwari2022limited}. However, these methods can only train their denoising models in the projection and image domain separately with independent loss functions or require paired data.\\
In this work, we present a self-supervised denoising pipeline that can be trained end-to-end starting from the raw projection data and predicting a denoised reconstruction. Our method allows integrating any trainable denoising model in the projection and the image domain (dual-domain) as illustrated in Fig.~\ref{fig:overview}. All operators are optimized simultaneously by backpropagating a gradient through the entire pipeline including the reconstruction operator without requiring high-dose target data. Together with our proposed denoising pipeline, we make our Python framework for loading, rebinning, and reconstructing all projection data directly from DICOM-CT-PD format \cite{chen2015development} publicly available to facilitate the usage of helical CT data. Our contributions are several-fold.
\begin{itemize}
  \item We present a dual-domain, end-to-end trainable, and entirely self-supervised CT denoising and reconstruction pipeline.
  \item We demonstrate the effectiveness of our dual-domain approach on both medical CT and pre-clinical X-ray Microscope (XRM) data starting from the raw acquired projection images.
  \item We make our helical projection rebinning and reconstruction framework publicly available to simplify using helical CT data and provide an open-source differentiable reconstruction pipeline for medical CT.
\end{itemize}

\section{Methods}
\label{sec:methods}

\subsection{End-to-end CT denoising}
\label{ssec:denoising}

During CT acquisitions image $\mathbf{y}$ is measured using the forward projection operator $\mathbf{A}$. Projection images $\mathbf{x}$ are generated that are affected by noise $\mathbf{n}$ through photon statistics and detector physics
\begin{align}
    \mathbf{x} = \mathbf{Ay} + \mathbf{n}\enspace.
\end{align}
A linear CT reconstruction operator $\mathbf{R}$, e.g., filtered back projection (FBP), can be used to reconstruct a noise-affected version of the measured image
\begin{align}
    \mathbf{y'} = \mathbf{Rx} =  \mathbf{RAy} + \mathbf{Rn}\enspace.
\end{align}
In the measurement domain, noise is a mixture of Poisson and Gaussian distributions defined through the acquired photon statistics and electric noise on the detector \cite{yu2012development}. The reconstruction operator acts on $\mathbf{x}$ and thereby distributes noise over the entire reconstructed image leading to a complex noise pattern. Therefore, already denoising in the measurement domain can be advantageous due to the relatively simple noise distribution present. Denoising operators $\mathbf{D}_{\text{proj}}^w$ and $\mathbf{D}_{\text{img}}^w$ dependent on trainable parameters $w$ can be used to remove noise at different stages of the reconstruction pipeline to predict a denoised image representation $\mathbf{\hat{y}}$
\begin{align}
    \mathbf{\hat{y}} = \mathbf{D}_{\text{img}}^w \mathbf{R} \mathbf{D}_{\text{proj}}^w \mathbf{x}\enspace.
\end{align}
Subsequently, a loss $L$ can be calculated as a quality measure of the prediction. To allow optimizing the set of trainable parameters $w_\text{proj}$ of $\mathbf{D}_{\text{proj}}^w$, the gradient
\begin{align}
    \frac{\partial L}{\partial w_\text{proj}} = \frac{\partial L}{\partial \mathbf{\hat{y}}}\frac{\partial \mathbf{\hat{y}}}{\partial w_\text{proj}}
    \label{eq:gradient_flow}
\end{align}
must be derived, which requires a differentiable reconstruction operator $\mathbf{R}$. In this work, we employ differentiable fan-beam \cite{ronchetti2020torchradon} and cone beam \cite{syben2019pyro} FBP operators that can backpropagate a loss in the measurement domain. Our combined end-to-end trainable pipeline with denoising operators in both domains is illustrated in Fig. \ref{fig:reco_pipeline}.

\subsection{Self-supervised training}
\label{ssec:noise2inverse}

The Noise2Inverse approach presents an image noise quality metric that does not require high-dose target data \cite{hendriksen2020noise2inverse}. The idea is to, first, split data into multiple element-wise independent sets, second, denoise a subsection of the sets, and third, calculate the distance of the prediction to the remaining sets. In CT applications data must be split in the measurement domain to preserve element-wise independence as the reconstruction operator $\mathbf{R}$ distributes each projected view over the entire image. In practice, we split our projection data into two independent sets $\mathbf{x_1}$ and $\mathbf{x_2}$ containing the projections with odd and even indices respectively. Subsequently, the projection sets are processed independently to obtain element-wise independent reconstructions $\mathbf{\hat{y}}_i$ and $\mathbf{\hat{y}}_j$
\begin{align}
    \begin{split}
        \mathbf{\hat{y}}_i &= \mathbf{D}_{\text{img}}^w \mathbf{R} \mathbf{D}_{\text{proj}}^w \mathbf{x}_i\\
        \mathbf{\hat{y}}_j &= \mathbf{R} \mathbf{x}_j
    \end{split}
\end{align}
with $i,j \in \{ 1,2 \}$ and $i\neq j$. Krull~\textit{et al.} \cite{hendriksen2020noise2inverse} proved that by minimizing the mean-squared error (MSE) between $\mathbf{\hat{y}}_i$ and $\mathbf{\hat{y}}_j$ denoising models learn to predict the underlying noise-free image. The final denoised prediction during inference is derived from all projections by averaging both denoised reconstructions
\begin{align}
    \mathbf{\hat{y}} = \frac{1}{2} \left(\mathbf{\hat{y}}_{i=1} + \mathbf{\hat{y}}_{i=2}\right)\enspace.
\end{align}
In our proposed dual-domain denoising pipeline we propagate this self-supervised loss back through the reconstruction operator to the projection denoising operator following the setting described in Sec.~\ref{ssec:denoising} and Fig.~\ref{fig:reco_pipeline}.

\subsection{Projection rebinning}
\label{ssec:rebinning}
Most medical CT scanners acquire projections on helical trajectories to reduce scan times and patient dose. However, to the best of our knowledge, there is no differentiable reconstruction operator available that supports helical acquisition geometries. Therefore, we rebinned helical projection data to fan-beam geometry following the algorithm of Noo~\textit{et al.} \cite{noo1999single} to enable backprojecting with differentiable operators as described in Sec.~\ref{ssec:denoising}. We made our repository publicly available that loads projection and geometry data from raw DICOM-CT-PD format \cite{chen2015development} used for all projections in the largest public low-dose CT data set \cite{moen2021low}, rebins the projections to fan-beam geometry, and reconstructs them using differentiable fan-beam FBP \cite{ronchetti2020torchradon}. We believe that our open-source Python framework can remove barriers for other researchers when developing algorithms for medical CT data \footnote{\url{https://github.com/faebstn96/helix2fan}}.

\section{Experiments}
\label{sec:experiments}
In this work, we perform multiple experiments on two distinct CT data sets to demonstrate the effectiveness of our proposed dual-domain, end-to-end trainable, self-supervised denoising pipeline. First, we perform experiments on rebinned helical abdomen CT scans ($25\,\%$ dose) to show applicability in a clinical setting. Second, we show that dose and acquisition speed can be improved in pre-clinical cone-beam X-ray microscope (XRM) scans on mouse bone samples ($10\,\%$ dose). Future in vivo XRM acquisitions of the bone-remodeling process on the micrometer scale can help to understand and develop treatments for bone-related diseases \cite{gruneboom2019next}. We used the differentiable cone-beam reconstruction pipeline by Thies~\textit{et al.} \cite{thies2022calibration} in our XRM experiments.\\
Three different denoising settings were investigated: (a) self-supervised denoising following Sec.~\ref{ssec:noise2inverse} and (b) supervised denoising, both using denoising operators in the projection and the image domain. In addition, we performed (c) self-supervised denoising with only one denoising operator as reconstruction post-processing as it is done in many related works including Noise2Inverse \cite{hendriksen2020noise2inverse}. We investigated the compatibility of two different denoising operators to our pipeline: First, standard U-Net architectures \cite{ronneberger2015u} which can be regarded as representative of most CNN-based methods. Second, single trainable bilateral filters (BFs) \cite{wagner2022ultra} which are conventional/hybrid ultralow-parameter (four trainable parameters) filters that have been shown to achieve competitive and robust denoising performance compared to deep neural networks \cite{wagner2022trainable}. Whereas BFs were directly employed to predict denoised images, the U-Nets were used to predict the residual noise from the network input, which was subsequently subtracted from that input. This setting turned out to converge more stably after the random weight initialization.\\
Both data sets were split into four training, one validation, and five test scans respectively with each scan reconstructed to either $100$ (abdomen CT) or $30$ (XRM) slices. We trained on single CT slices due to limited GPU size (Nvidia RTX A6000) but tested on the entire scans. The training and validation data are only used during supervised training. We used the Adam optimizer with lr $= 5 \cdot 10^{-5}$ (U-Net) and $5 \cdot 10^{-3}$ (BFs) in all our experiments and trained until convergence of the self-supervised loss (experiment (a) and (c)) or validation loss (experiment (b)).

\section{Results and discussion}
\label{sec:results}
\begin{figure}[h!tb]
\begin{minipage}[b]{1.0\linewidth}
  \centering
  \centerline{\includegraphics[width=8.5cm]{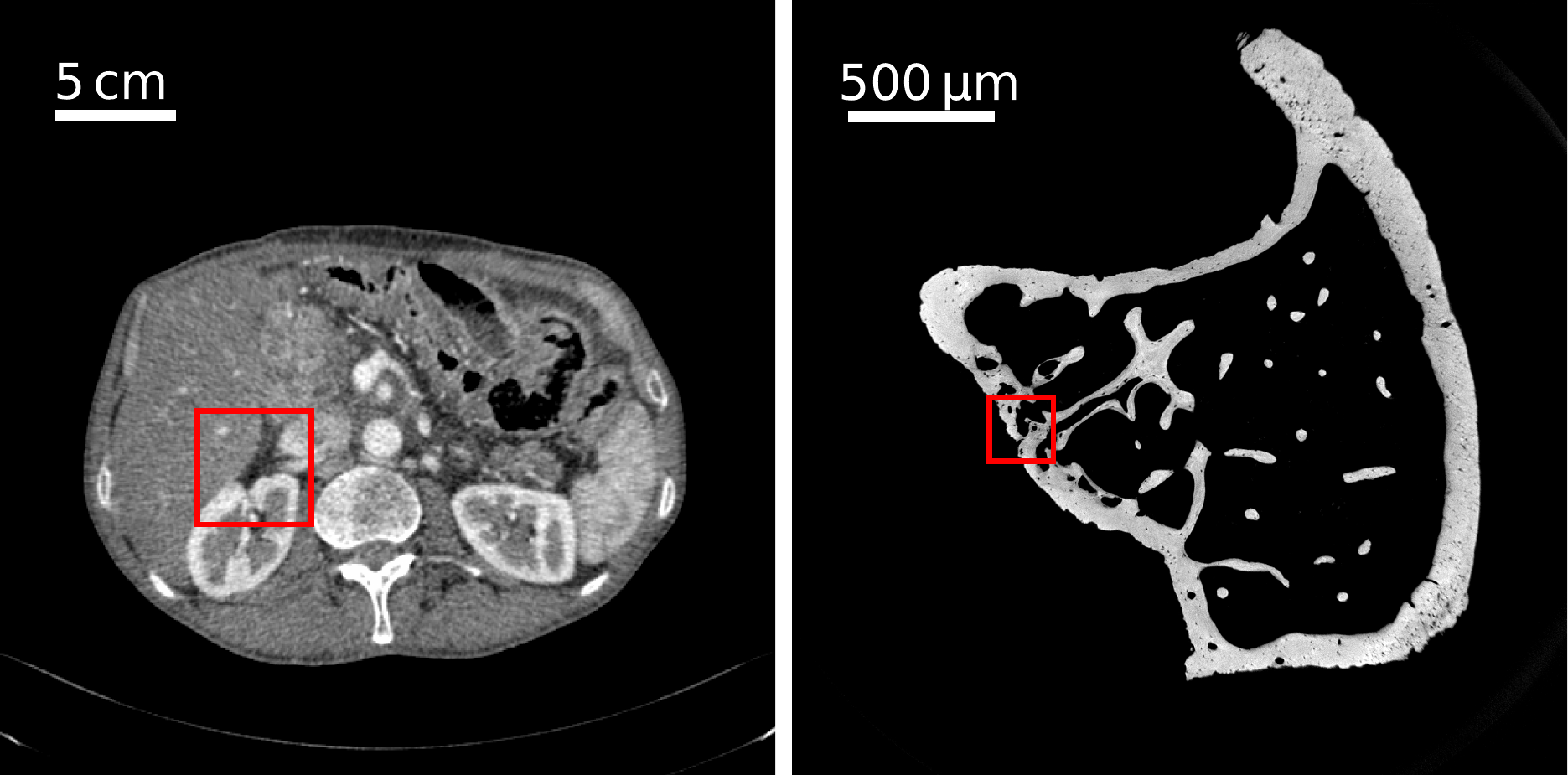}}
  \vspace{0.05cm}
\end{minipage}
\begin{minipage}[b]{1.0\linewidth}
  \centering
  \centerline{\includegraphics[width=8.5cm]{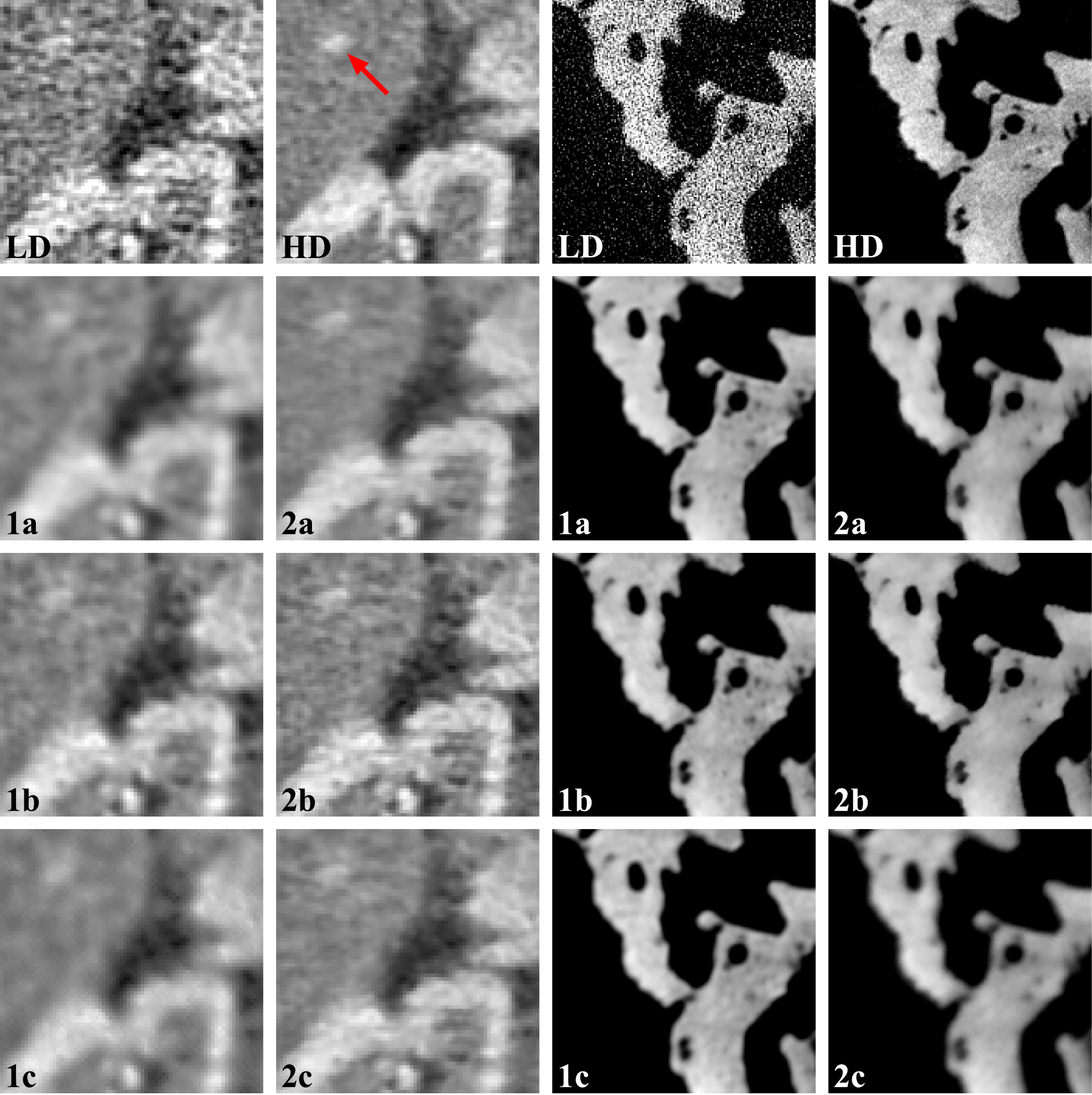}}
\end{minipage}
\caption{Examplary predictions of an abdomen CT slice (left) and the cross sections of a mouse tibia bone (XRM) close to the knee area (right). Below high-dose overview images, ROIs (red squares) of (LD) low-dose, (HD) high-dose, (1a) dual U-Nets (self-supervised), (1b) dual U-Nets (supervised), (1c) reco U-Net (self-supervised), (2a) dual BFs (self-supervised), (2b) dual BFs (supervised), and (2c) reco BF (self-supervised) predictions are presented. Windows are $[-150, 250]\,\text{HU}$ (abdomen) and $[0.05, 0.32]\,\text{arb.}\,\text{unit}$ (XRM).}
\label{fig:vis_res}
\end{figure}

We present quantitative and qualitative results for all three investigated training strategies (a), (b), and (c). Quantitative quality measures peak signal-to-noise ratio (PSNR) and structural similarity index (SSIM) are calculated across all test scans (mean $\pm$ std) and listed in Tab.~\ref{tab:abdomen} and Tab.~\ref{tab:xrm} for the investigated abdomen and bone scans. 
\begin{table}[ht]
\caption{Quantitative results on abdomen CT data. Dual-domain self-supervised (a) and supervised (b) as well as solely post-processing (c) denoising is investigated. (1) indicates the U-Net and (2) the BFs.}
\begin{tabular}{lcc}
\hline
& PSNR & SSIM \\
\hline
Low-dose & $41.7 \pm 1.4$ & $0.941 \pm 0.017$ \\
\hline
(1a) Dual U-Nets (self-sup) & $44.8 \pm 1.2$ & $0.975 \pm 0.006$ \\
(1b) Dual U-Nets (sup) & $45.6 \pm 1.2$ & $0.978 \pm 0.005$ \\
(1c) Reco U-Net (self-sup) & $43.4 \pm 1.0$ & $0.965 \pm 0.007$ \\
\hline
(2a) Dual BFs (self-sup) & $45.0 \pm 1.4$ & $0.977 \pm 0.006$ \\
(2b) Dual BFs (sup) & $45.3 \pm 1.4$ & $0.976 \pm 0.008$ \\
(2c) Reco BF (self-sup) & $44.1 \pm 1.1$ & $0.973 \pm 0.006$ \\
\hline
\end{tabular}
\label{tab:abdomen}
\end{table}\noindent
In general, across both data sets and both model types (U-Nets, BFs), supervised training using the ground-truth high-dose reconstructions to train the network outperformed self-supervised methods quantitatively by a small but distinct amount. In addition, all self-supervised pipelines using denoising operators in both projection and image domain outperformed the respective self-supervised model that only performs image post-processing by $82.4\text{--}94.1\,\%$/$12.5\text{--}41.7\,\%$ (PSNR/SSIM) on abdomen data and by $1.5\text{--}2.9\,\%$/$0.4\text{--}0.5\,\%$ (PSNR/SSIM) on XRM scans relative to the low-dose baseline. Therefore, we conclude that dual-domain CT denoising is beneficial over single-domain denoising.\\
\begin{table}[t]
\caption{Quantitative results on XRM data. Abbreviations as in Tab.~\ref{tab:abdomen}.}
\begin{tabular}{lcc}
\hline
& PSNR & SSIM \\
\hline
Low-dose & $18.6 \pm 0.1$ & $0.141 \pm 0.009$ \\
\hline
(1a) Dual U-Nets (self-sup) & $32.5 \pm 0.1$ & $0.671 \pm 0.008$ \\
(1b) Dual U-Nets (sup) & $32.7 \pm 0.1$ & $0.682 \pm 0.008$ \\
(1c) Reco U-Net (self-sup) & $32.3 \pm 0.1$ & $0.668 \pm 0.008$ \\
\hline
(2a) Dual BFs (self-sup) & $32.6 \pm 0.1$ & $0.686 \pm 0.008$ \\
(2b) Dual BFs (sup) & $32.9 \pm 0.1$ & $0.690 \pm 0.008$ \\
(2c) Reco BF (self-sup) & $32.2 \pm 0.2$ & $0.684 \pm 0.009$ \\
\hline
\end{tabular}
\label{tab:xrm}
\end{table}\noindent
In general, the experiments on medical data show a stronger relative improvement with respect to the low-dose baseline. However, different image content, data ranges, and noise levels in the two investigated data sets make quality metrics and improvements difficult to compare. In addition, we believe that due to the high angular sampling in XRM scans that data inherently contains fewer reconstruction artifacts, which can simplify denoising during post-processing.\\
Magnified ROIs of model predictions on both data sets are presented in Fig.~\ref{fig:vis_res}. A liver lesion is highlighted for the abdomen CT data (red arrow). Likewise to the quantitative results, the supervisedly trained models (1b, 2b) predict reconstructions closest to the high-dose ground-truth images. The dual-domain models employing denoising operators in both the projection and image domain simultaneously (1a, 2a) reduce noise compared to the noisy low-dose image and outperform the respective model only using a post-processing denoising operator (1c, 2c).\\
In general, our experiments show that our presented dual-domain and self-supervised CT denoising pipeline improves denoising compared to pure reconstruction post-processing. The benefit of projection denoising can be explained through the distinct noise distribution in the projection data, which constitutes a considerably easier denoising task compared to complex noise removal on the reconstruction. We hope that our open-source projection rebinning and differentiable reconstruction framework can facilitate more research on methods intervening in the different data domains of CT reconstruction pipelines.

\section{Conclusion}
\label{sec:conclusion}
In this work, we presented an end-to-end trainable and self-supervised CT reconstruction pipeline that performs denoising in two domains, namely projection and image domain. Our experiments on medical and pre-clinical CT data demonstrate quantitatively and qualitatively that dual-domain denoising is beneficial over solely reconstruction image denoising as conducted in many recent works. We believe that our released open-source helical CT rebinning and differentiable reconstruction framework can enable further research on self-supervised and dual-domain CT pipelines.

\newpage
\section{Compliance with ethical standards}
\label{sec:ethics}

The abdomen CT study was conducted retrospectively using human subject data made available in open access by Moen~\textit{et al.} \cite{moen2021low}. Ethical approval was not required as confirmed by the license attached with the open-access data. The bone XRM study was performed in line with the principles of the Declaration of Helsinki. Approval was granted by the Ethics Committee of FAU Erlangen-Nürnberg (license TS-10/2017).

\section{Acknowledgments}
\label{sec:acknowledgments}

This work was supported by the European Research Council (ERC Grant No. 810316) and a GPU donation through the NVIDIA Hardware Grant Program. F.W. conceived and conducted the experiments. M.T., L.P., N.M., M.R., M.G., J.U., and F.D. provided valuable technical feedback during development. O.A., S.P., and D.W. prepared and scanned the bone samples. A.M. supervised the project. All authors reviewed the manuscript. L.P., N.M., M.R., and F.D. are employees of Siemens Healthcare GmbH.

\bibliographystyle{IEEEbib}
\bibliography{refs}

\end{document}